\documentclass[review,showpacs,amsmath,amstex,amssymb, mathfonts]{elsarticle}
\pdfoutput=1
\usepackage{amsthm,color,amsfonts,graphicx,verbatim}
\usepackage{amsmath}
\usepackage{amssymb}
\usepackage{amsthm}
\usepackage{amsfonts}
\usepackage{listings}
\usepackage{enumerate}
\usepackage{latexsym}
\usepackage{psfrag}
\usepackage{bm}
\usepackage{graphicx}

\usepackage{hyperref}
\hypersetup{
    bookmarks=true,         
    unicode=false,          
    pdftoolbar=true,        
    pdfmenubar=true,        
    pdffitwindow=false,     
    pdfstartview={FitH},    
    pdftitle={Periodically driven ergodic and many-body localized quantum systems},    
    pdfauthor={},     
    pdfsubject={},   
    pdfcreator={},   
    pdfproducer={}, 
    pdfkeywords={keyword1} {key2} {key3}, 
    pdfnewwindow=true,      
    colorlinks=true,       
    linkcolor=black,          
    citecolor=blue,        
    filecolor=magenta,      
    urlcolor=blue           
}

\newcommand{\be}{\begin{equation}}
\newcommand{\ee}{\end{equation}}
\newcommand{\bea}{\begin{eqnarray}}
\newcommand{\eea}{\end{eqnarray}}

\newcommand{\la}{\langle}
\newcommand{\ra}{\rangle}

\renewcommand{\phi}{\varphi}
\renewcommand{\epsilon}{\varepsilon}

\begin{document}

\begin{frontmatter}

\title{Periodically driven ergodic and many-body localized quantum systems}

\author{Pedro Ponte,$^{1,2}$ Anushya Chandran,$^1$ Z. Papi\'c,$^{1,3}$ and Dmitry A. Abanin$^{1,3}$}

\address{$^1$ Perimeter Institute for Theoretical Physics, Waterloo, ON N2L 2Y5, Canada}
\address{$^2$ Department of Physics and Astronomy, University of Waterloo, Ontario, N2L 3G1, Canada}
\address{$^3$ Institute for Quantum Computing, Waterloo, ON N2L 3G1, Canada}

\date{\today}

\begin{abstract}
We study dynamics of isolated quantum many-body systems whose Hamiltonian is switched between two different operators periodically in time. The eigenvalue problem of the associated Floquet operator maps onto an effective hopping problem. Using the effective model, we establish conditions on the spectral properties of the two Hamiltonians for the system to localize in energy space. We find that ergodic systems always delocalize in energy space and heat up to infinite temperature, for both local and global driving. In contrast, many-body localized systems with quenched disorder remain localized at finite energy. We support our conclusions by numerical simulations of disordered spin chains.
We argue that our results hold for general driving protocols, and discuss their experimental implications. 
\end{abstract}



\end{frontmatter}

\section{Introduction}

Quantum systems coupled to time-varying external fields are ubiquitous in nature. 
They exhibit many interesting phenomena including the laser, the maser, electron-spin resonance and nuclear magnetic resonance (NMR)~\cite{Aranson:2002aa,Grifoni:1998cs}. 
The experimental developments in ultra-cold atomic or molecular gases and trapped ions in the last two decades have taken us beyond the few-atom systems into the regime of isolated \emph{interacting} systems, whose quantum dynamics reveals novel aspects of thermalization, transport and non-linear response~\cite{Polkovnikov:2011ys,Bloch:2008ly}.

Periodically driven systems can exhibit non-trivial steady states, even in the non-interacting limit \cite{Casati:1979xq, Grempel:1984ye,Moore:1994wt,Lemarie:2009ee,Russomanno:2012qe}.
An illustrative system is the kicked quantum rotor, which can dynamically localize in momentum space \cite{Casati:1979xq, Grempel:1984ye,Moore:1994wt,Lemarie:2009ee}. 
Periodic driving can also be used to control the band structure and induce topological states ~\cite{Kitagawa:2010aa,Kitagawa:2011aa,Lindner:2011aa,Wang:2013aa}.

Here we study periodically driven {\it many-body} systems with local interactions. This problem has recently been addressed by D'Alessio and Polkovnikov~\cite{DAlessio:2013rm} who hypothesized that two distinct dynamical regimes are possible: the system either (i) keeps absorbing energy, heating up to infinite temperature (e.g., defined using the time-averaged Hamiltonian) at long times, or (ii) dynamically localizes at a certain energy, similar to the case of the kicked rotor.

The long-time behaviour of a driven system is determined by the properties of the so-called Floquet Hamiltonian $\hat{H}_F$, defined in terms of the Floquet operator 
\begin{equation}
\hat{F}=e^{-i\hat{H}_FT},
\end{equation}
which is the evolution operator over one period, 
\begin{equation}
\hat F={\mathcal T}\int_{0}^T \exp(-i\hat H(t)t)\, dt.
\end{equation}
Here $\hat H(t+T)=\hat H$ is the system's Hamiltonian, and $\mathcal T$ is time-ordering. The Floquet Hamiltonian, which determines the time evolution of the system, can be calculated perturbatively in the driving period $T$ using the Magnus expansion~\cite{Magnus}. The convergence of the Magnus expansion implies that there exists a physical $\hat{H}_F$ that is a sum of local terms, such that the driving dynamics is equivalent to a single quench of the Hamiltonian~\cite{DAlessio:2013rm}. 
In this case, the system retains the memory about $\hat{H}_F$ and is at finite temperature with respect to $\hat{H}_F$ after many periods. On the contrary, if the system heats up to infinite temperature at long times, the Magnus expansion does not converge.

\begin{figure}[htb]
\begin{center}
\includegraphics[width=\columnwidth]{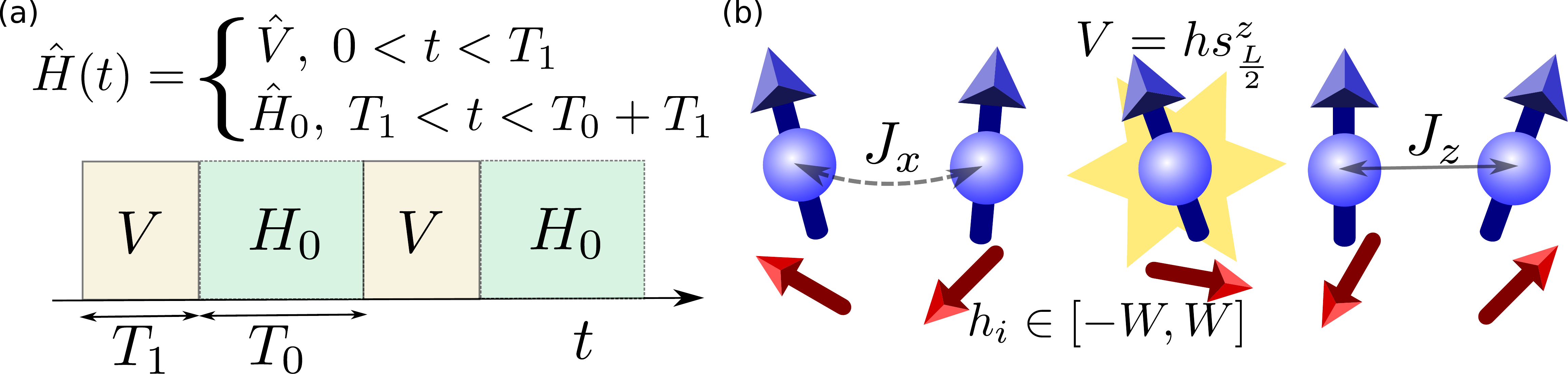}
\caption{\label{Fig:model} (Color online) (a) A scheme of the general driving protocol. (b) Example of a 1d XXZ spin chain studied numerically (blue arrows). The Hamiltonian $H_0$ contains nearest-neighbor hopping and interactions, in the presence of random $z$-field (red arrows). Driving is performed locally by the local operator $\hat{V}=hs_{L/2}^z$ applied to the middle spin.
}
\end{center}
\end{figure}
Here we establish the conditions under which the two dynamical regimes are realized in locally driven many-body systems. We consider a driving protocol illustrated in Fig.~\ref{Fig:model}(a) where the Hamiltonian is switched between two operators periodically in time,
\begin{align}
\label{Eq:Hswitch}
\hat{H}(t) = \begin{cases} \hat{V} & 0<t<T_1 \\
\hat{H}_0 & T_1<t<T_0+T_1
\end{cases}
\end{align} 
i.e., $\hat{H}_0$($\hat{V}$) is applied during time $T_0$ ($T_1$), such that the total period $T=T_0+T_1$. We will consider lattice systems, in which the Hilbert space on every site is finite-dimensional (e.g., an interacting spin system).

In Section~\ref{sec:mapping} we map the spectral problem for the Floquet operator describing this system onto an effective hopping problem, and show that the competition between the typical matrix elements of $\tan(\hat{V} T_1/2)$ between the eigenstates of $H_0$ and the typical energy spacings determines whether or not the system will heat up to infinite temperature at long times. In Section~\ref{sec:numerics} we discuss some of the signatures of heating in the hopping problem, and introduce the spin model that is studied numerically. In Sections~\ref{sec:ergodic},\ref{sec:mbl} we then apply our criterion to two general classes of driven many-body systems, considering \emph{local} $\hat{V}$ that acts on a few nearby degrees of freedom. The first class is ergodic systems -- i.e., systems that act as their own heat bath and satisfy the eigenstate thermalization hypothesis (ETH)~\cite{Deutsch:1991ss,Srednicki:1994dw,Rigol:2008bh}.
In Section~\ref{sec:ergodic} we show that for ergodic systems, the system heats up to infinite temperature under driving, and thus the Magnus expansion in the driving period $T$ diverges, and $\hat{H}_F$ is unphysical.
The second class is {\it many-body localized} (MBL) systems with quenched disorder that are known to be non-ergodic \cite{Basko:2006aa, Pal:2010gs, Oganesyan:2007aa, Serbyn:2013rt, Huse:2013kq}. 
Under local driving, we show that MBL systems retain memory of their initial state and never reach infinite temperature (Section~\ref{sec:mbl}). Here, the Magnus expansion converges, $\hat{H}_F$ is local, and itself MBL.
Throughout Sec.~\ref{sec:localdriving} we support our analytical results with numerical studies of the driven XXZ spin-$1/2$ chain in random $z$-fields,  Fig.~\ref{Fig:model}(b). Our conclusions are summarized in Section~\ref{sec:conclusions}. 

The above results indicate that driven interacting systems differ from non-interacting systems (on a lattice): the latter do have a set of conserved quantities, a local Floquet Hamiltonian, and their long-time behaviour is described by the generalized Gibbs ensemble, rather than an infinite-temperature state~\cite{Lazarides:2014}. We note that periodically kicked random matrices, generally, do not show dynamical localization~\cite{Basko:2003}, sharing this property with one- and higher-dimensional ergodic systems.

\section{Mapping the Floquet onto a hopping problem}\label{sec:mapping}

In this Section we provide a mapping of the Floquet problem onto an effective hopping model. This mapping shows that  
the competition between the typical level spacing and the hopping matrix element (i.e., the matrix element of $\tan(\hat{V} T_1/2)$ between the eigenstates of $H_0$) determines the structure of the Floquet eigenstates.

The Floquet operator for the driving protocol in Fig.~\ref{Fig:model}(a) is given by 
\begin{equation}
\hat{F}=\exp(-i\hat{H}_0T_0)\exp(-i\hat{V}T_1).
\end{equation}
The eigenstates of $\hat{F}$ completely determine the stroboscopic evolution of the system. Below, we map the eigenvalue problem of  $\hat{F}$ onto a hopping problem, similar to the kicked rotor model in Ref.~\cite{Grempel:1984ye}. The lattice sites of the hopping problem represent eigenstates of $\hat{H}_0$, while $\hat{V}$ induces hopping between sites.   

The Floquet operator is unitary. Its spectrum is: 
\be\label{eq:eigenvalues}
\hat{F}|\psi_i\ra=e^{-i \hat{H}_F T}|\psi_i\ra=e^{-i\omega_i T}|\psi_i\ra, \,\, i=1,\ldots,{\mathcal M}, 
\ee
where ${\mathcal M}$ is the dimensionality of the Hilbert space (e.g., $\mathcal{M} = 2^N$ for the system of $N$ $1/2$-spins considered below) and $\la \psi_i | \psi_j \ra = \delta_{ij}$. The quasi-energies $\omega_i$ are defined modulo $2\pi/T$, hence $\hat{H}_F$  is not unique. 

Finding the spectrum of $\hat{F}$ in a many-body system is generally hard, due to $\hat F$ being highly non-local. 
To circumvent this difficulty, let us provide an explicit mapping to a local Hamiltonian problem.
Rewrite $e^{-i\hat VT_1}$ in terms of a Hermitian operator $\hat{G}$ as:
\be\label{eq:V_rewrite}
e^{-i\hat{V}T_1}=(1+i\hat{G})(1-i\hat{G})^{-1}, \,\, \hat{G}= -\tan (\hat{V}T_1/2).
\ee
In general $\hat{G}$ is not spatially local.
If however $\hat{V}$ is local (that is, acts non-trivially only on a finite number of spatial degrees of freedom in thermodynamic limit), then $\hat{G}$ is also local. Defining $|\chi_i\ra \equiv (1-i\hat{G})^{-1}|\psi _i\ra$, Eq.~\eqref{eq:eigenvalues} becomes: 
\be\label{eq:mapping2}
\left[ \tan \frac{1}{2}(\hat{H}_0 T_0 - \omega_i T)  - \hat{G} \right] |\chi_i\ra=0.
\ee
Let us view the eigenbasis of $\hat H_0$, labeled by $|\alpha\ra$, as sites in a lattice. Solving Eq.~\eqref{eq:mapping2} is equivalent to finding the zero-energy eigenstate of a hopping problem on this lattice, where $\tan\frac{E_\alpha T_0-\omega_iT}{2}$ plays the role of an on-site energy on site $\alpha$, and $G_{\alpha\beta}$ is the hopping amplitude between sites $\alpha$ and $\beta$. The competition between the typical level spacing and the hopping matrix element, which is different in the ergodic and MBL phases \cite{Srednicki:1999bd,Khatami:2013et,Beugeling:2013zh,Pal:2010gs}, thus determines the structure of the Floquet eigenstates. 

\section{Microscopic model and heating diagnostics}\label{sec:numerics}

In this Section we define several quantities that can serve as a measure for when the system heats up under periodic driving. We also introduce the model of a disordered XXZ spin chain, where these quantities can be readily computed in numerical simulations. These results are presented in Sec.~\ref{sec:localdriving}.

We imagine preparing the system at $t=0$ in a low-energy eigenstate of $\hat{H}_0$, $|\phi_0\ra=|\alpha_0\ra$. The stroboscopic evolution at times $t=N T$ follows from the expansion of $|\phi_0\ra$ in the Floquet eigenbasis:
\begin{equation}
|\phi_N\ra=\hat{F}^N|\phi_0\ra=\sum _i A_{\alpha_0 i} \exp(-i\omega_i NT)|\psi_i\ra,
\end{equation}  
where $A_{\alpha i}=\la \psi_i| \alpha  \ra$. At long times, the time-averaged density matrix is 
\begin{equation}
\hat\rho_{\infty}=\sum_{i} |A_{\alpha_0 i}|^2 |\psi_i\ra\la \psi_i|.
\end{equation}
The nature of the eigenstates $|\psi_i\ra$ determines the steady state as $t\rightarrow\infty$. If each $|\psi_i\ra$ is delocalized in the eigenbasis of $\hat{H}_0$, then each $|\psi_i\ra$ corresponds to an infinite temperature state. The entire density matrix, $\hat\rho_{\infty}$, describes a system at infinite temperature in this case. If on the other hand the $|\psi_i\ra$ are localized in the eigenbasis of $\hat{H}_0$, then, depending on $A_{\alpha_0 i}$, $\hat\rho_{\infty}$ describes a system at different energies. 

To characterize the energy absorbed under driving, at each $t=NT$ we introduce a dimensionless energy
\begin{equation}\label{eq:QN}
Q_N= \frac{(\la \psi_N|\hat H_0|\psi_N\ra-E_{0})}{(E_{T=\infty}-E_0)},
\end{equation}
where $E_{T=\infty}={\rm Tr}(H_0)/{\mathcal M}$ is the average energy at infinite temperature, and $E_0$ is the energy at $t=0$. Ergodic and MBL cases are distinguished by $Q_N \rightarrow 1$ and $Q_N \rightarrow 0$ as $N\rightarrow \infty$, respectively.

Finally, to quantify the structure of the Floquet eigenstates, we compute the participation ratio (PR). For the Floquet eigenstate $|\psi_i\ra$, PR in the basis of eigenstates $\alpha$ is defined as 
\begin{equation}\label{eq:pr}
\textrm{PR} =  \frac{1}{\mathcal{M}} \left( \sum_\alpha  |A_{\alpha i}|^4 \right)^{-1}.
\end{equation}

The above quantities can be readily computed in finite spin chains using exact diagonalization. For concreteness, 
we focus on the XXZ spin-$1/2$ chain with $L$ sites and open boundary conditions, illustrated in Fig.~\ref{Fig:model}: 
\be\label{eq:H0}
\hat H_0=J_x\sum_{i} (s_{i}^x s_{i+1}^x+s_{i}^y s_{i+1}^y)+J_z\sum_{i} s_{i}^z s_{i+1}^z+\sum_i h_i s_{i}^z, 
\ee
where fields $h_i$ are independent random variables drawn from the uniform distribution $[-W,W]$, and we fix $J_x=J_z=1$. 
The model (\ref{eq:H0}) exhibits both ergodic and MBL phases as a function of disorder strength $W$, with the transition at $W_*\approx 3$~\cite{Pal:2010gs}. The system is driven by
\begin{equation}\label{eq:V}
V=h s_{L/2}^z
\end{equation}
acting on the middle spin (we assume $h=2$). In order to minimize finite-size effects and thus satisfy $T_0 \Delta \gg 1$, where $\Delta$ is the bandwidth, we fix $T_0=7$. The results for this model will be presented in Sec.~\ref{sec:localdriving}, for both the ergodic and MBL cases which correspond to disorder strength $W=0.5$ and $W=8$, respectively. Number of disorder averages performed ranges from $2\times 10^4$ ($L=8,10$) to $\sim 10^3$ ($L=12, 14$).

\section{Local driving}\label{sec:localdriving}

In this Section we derive the conditions for the system to heat up to infinite temperature, and confirm them numerically.
We will assume that $\hat V$ is local (e.g., an operator of the form in Eq.~(\ref{eq:V})). Then,  $\hat G=-\tan \frac{\hat{V}T_1}{2}$ is also local. Furthermore we will assume that the Hilbert space at every lattice site is finite-dimensional, like in the model (\ref{eq:H0}), such that $\hat G$ has a finite operator norm, $||\hat G||\leq C$, except at special values of $T_1$ when $\eta_i T_1= (2n+1)\pi$, where $n$ is an integer and $\eta_i$ are the eigenvalues of $\hat V$. 

We expect the zero-energy states $|\chi_i\ra$ in Eq.~\eqref{eq:mapping2} to be localized in the eigenbasis of $\hat{H}_0$ when the typical hopping amplitude $G_{\alpha\beta}$ is much smaller than the level spacing $\Delta \lambda$.  
At quasi-energy $\omega$, the on-site energies $\lambda_\alpha(\omega)=\tan\frac{E_\alpha T_0-\omega T}{2}$ have a distribution $P(\lambda)\propto \frac{1}{1+\lambda^2}$ if the $E_\alpha$ are distributed uniformly. 
As we are searching for a zero-energy solution, we focus on $|\lambda_\alpha(\omega)|\lesssim 1$.
From the form of $P(\lambda)$, we see that the level spacing for $|\lambda_\alpha(\omega)|\lesssim 1$ is $\Delta \lambda \approx \Delta E \sim \frac 1{\mathcal M}$, where $\Delta E$ is the level spacing in energies of $\hat{H}_0$. 
The criterion for localization thus takes the form:
\be\label{eq:criterion}
|G_{\alpha\beta}|=|\la \alpha |\hat G|\beta\ra|\ll 1/\mathcal M. 
\ee
Two comments are in order. First, the sites that we have not included have energy $\lambda_\alpha(\omega) \gg 1$, as well as level spacing $\Delta \lambda \gg 1$. Such states form a small fraction of the Hilbert space in the thermodynamic limit and are extremely off-resonant; we are thus well-justified in ignoring them. 
Second, using Eq.~\eqref{eq:mapping2}, the Floquet eigenstates $|\psi_i\ra$ are localized (delocalized) if $|\chi_i\ra$ are localized (delocalized).

\begin{figure}[ttt]
\begin{center}
\includegraphics[width=\columnwidth]{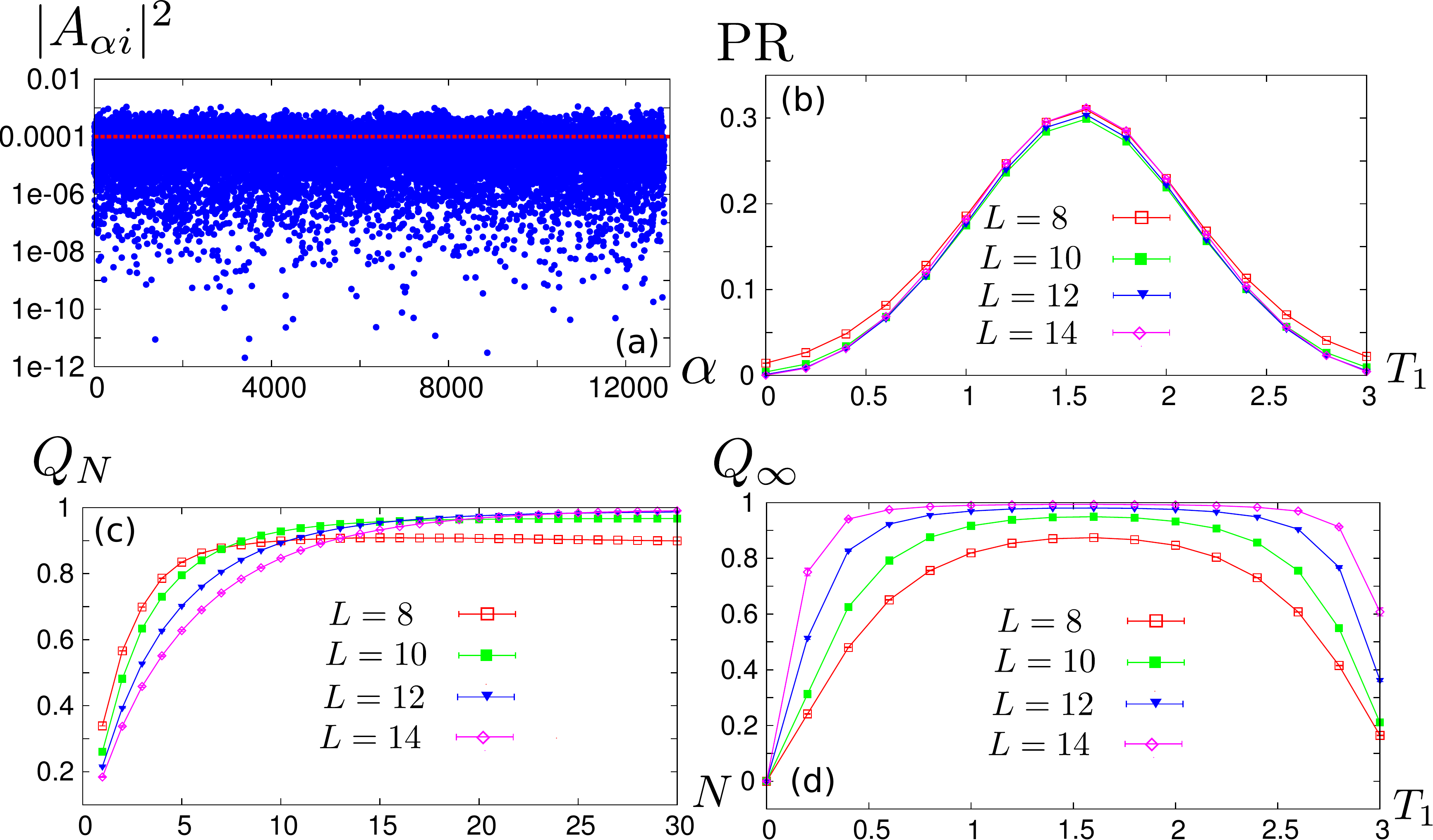}
\caption{\label{Fig:ergodic} (Color online) Energy delocalization and heating to infinite temperature in the ergodic phase of the XXZ model ($T_0=7, W=0.5$). (a) Squared overlap $|A_{\alpha i}|^2$ of the Floquet eigenstate $|\psi_i\ra$ with eigenstates $|\alpha\ra$ of $\hat H_0$, ordered by energy, for a fixed disorder realization and $T_1=1.5$. $|A_{\alpha i}|^2$ are nearly uniformly spread over all eigenstates $\alpha$. (b) Disorder-averaged PR vs. $T_1$. As $L$ is increased, PR remains finite in the ergodic phase.  (c) Disorder-averaged $Q_N$ vs. the number of driving cycles $N$ ($T_1=1.5$), for evolution starting from the ground state. (d) Disorder-averaged saturation value $Q_\infty$ vs. $T_1$ for different $L$. In the ergodic case, $Q_\infty$ sharply approaches $1$ as $L\to\infty$ for any $T_1$, signaling generic heating to infinite temperature. Number of disorder averages is $2\times 10^4$ ($L=8,10$) and $\sim 10^3$ ($L=12, 14$).
}
\end{center}
\end{figure}

\subsection{Ergodic systems}\label{sec:ergodic}

The condition (\ref{eq:criterion}) does not hold for ergodic systems even for arbitrarily weak driving. To see why, note that eigenstates $|\alpha\ra$ in the ergodic phase are like random vectors in the basis of product states that satisfy the ETH. The typical matrix element of {\it any} local operator between two random vectors is: 
\be\label{eq:Gerg}
|G_{\alpha\beta}|\sim  ||G ||/\sqrt{\mathcal M} \gg 1/\mathcal M.  
\ee
The scaling with the size of the Hilbert space is independent on the driving strength $||\hat G||$. Thus, in the thermodynamic limit, the hopping problem is always in the delocalized phase.

In the delocalized phase, the Floquet eigenstates are superpositions of nearly all eigenstates $|\alpha\ra$ with macroscopically different physical energies $E_\alpha$ (but with $\lambda_\alpha (\omega)\lesssim 1$). Their PR, Eq.~(\ref{eq:pr}), remains finite as $\mathcal{M}\rightarrow \infty$. The fraction of infinite-temperature states in the Hilbert space also increases as $\mathcal{M}\rightarrow \infty$; thus individual Floquet eigenstates describe infinite-temperature states of the system. 

We have verified the above statements numerically in the case of XXZ model, Eq.~(\ref{eq:H0}). In Fig.~\ref{Fig:ergodic}(a) we first illustrate the structure of a typical Floquet eigenstate $|\psi_i\ra$ for a fixed disorder realization and $T_1=1.5$. We plot $|A_{\alpha i}|^2$ as function of $\alpha$, ordered by energy $E_\alpha$. As expected for the ergodic case, a typical Floquet eigenstate is delocalized and has non-zero overlap with states at very different energies. 

The difference between Floquet eigenstates is further revealed in the behaviour of PR, Eq.~(\ref{eq:pr}), shown in Fig.~\ref{Fig:ergodic}(b). Disorder-averaged PR, plotted as a function of $T_1$ for different system sizes $L$, shows that Floquet eigenstates occupy a finite fraction of the Hilbert space in the thermodynamic limit when the system is ergodic. 

The energy absorbed after $N$ cycles, $Q_N$, Eq.~(\ref{eq:QN}), when the system is initially prepared in the ground state of $H_0$, is shown in Fig.~\ref{Fig:ergodic}(c). In the thermodynamic limit, $Q_N$ approaches 1 for ergodic systems~\footnote{We note that, since we consider finite systems, the system will undergo rare quantum revivals, when its energy becomes close to the initial value, which is much lower than the infinite-temperature value ($Q_N<1$). However, the revival time increases exponentially with system size, and therefore for all practical purposes revivals can be ignored.}. 

Similarly, the saturated value $Q_{\infty}$ in the ergodic phase (Fig.~\ref{Fig:ergodic}(d)) tends to 1 as $L$ is increased. Note that, for the system sizes studied here, $Q_{\infty} \ll 1$ for small $T_1$, due to the small norm of the operator $G$. However, $Q_{\infty}$ monotonically increases as a function of $L$, and is likely to reach 1 in the thermodynamic limit, even for for arbitrarily small values of $T_1$, suggesting that ergodic systems generally heat up to infinite temperature. 

\subsection{Many-body localized systems}\label{sec:mbl}

The situation is very different for MBL states. The typical matrix elements of local operators in the MBL phase decay exponentially with system size, but fall off faster than the level spacing \cite{Serbyn:aa}:
\be\label{eq:Gmbl}
G_{\alpha\beta}\sim ||G|| e^{-L/\xi}\ll 1/\mathcal M, 
\ee
where $\xi$ is the ``many-body" localization length (possibly different from the localization length of single-particle operators). The MBL phase is characterized by local integrals of motion with exponentially decaying tails~\cite{Serbyn:2013rt,Huse:2013kq}. Eigenstates $\alpha, \beta$ that have nearly the same energy typically differ by the values of the local integrals of motion a distance $\sim L$ away from the support of local operator $G$. The probability of changing the value of a remote integral of motion decays exponentially with distance, which explains the result (\ref{eq:Gmbl}). 

The criterion for localization (\ref{eq:criterion}) is satisfied in the MBL phase. Hopping $\hat G$ only significantly mixes a few eigenstates with a similar structure away from the support of $\hat G$ (that is, with the same values of the local integrals of motion distance $x\gtrsim \xi$ away). This implies that energy can be absorbed only in the vicinity of the driving. 

\begin{figure}[ttt]
\begin{center}
\includegraphics[width=\columnwidth]{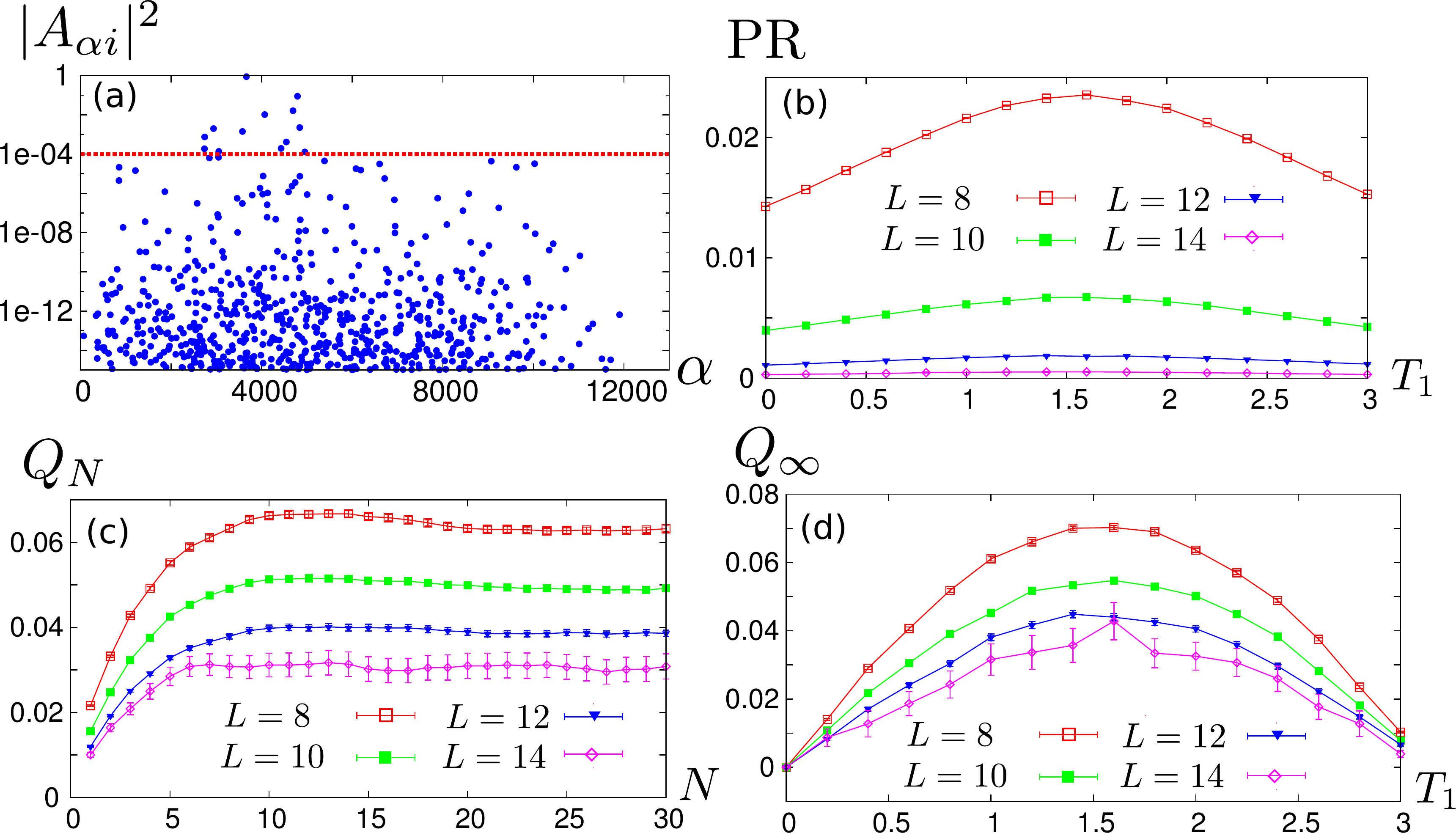}
\caption{\label{Fig:mbl} (Color online) Energy localization and absence of heating in the MBL phase ($T_0=7, W=8$). 
(a) Squared overlap $|A_{\alpha i}|^2$ of the Floquet eigenstate $|\psi_i\ra$ with eigenstates $|\alpha\ra$ of $\hat H_0$, ordered by energy, for a fixed disorder realization and $T_1=1.5$. The overlap is non-zero only for a few eigenstates with similar energies.
(b) Disorder-averaged PR vs. $T_1$. As $L$ is increased, PR decreases as $1/\mathcal{M}$. 
(c) Disorder-averaged $Q_N$ vs. the number of driving cycles $N$ ($T_1=1.5$), for evolution starting from the ground state.
(d) Disorder-averaged saturation value $Q_\infty$ vs. $T_1$ for different $L$. In the MBL phase, $Q_\infty \ll 1$ and decreases with $L$, indicating that the system absorbs finite energy locally. 
Number of disorder averages is $2\times 10^4$ ($L=8,10$) and $\sim 10^3$ ($L=12, 14$).
}
\end{center}
\end{figure}

The structure of  the Floquet eigenstates in the MBL phase is thus very different from the ergodic case. A typical Floquet eigenstate in the MBL case has sizable overlap only with those $|\alpha\rangle$ that are close in energy, as shown in Fig.~\ref{Fig:mbl}(a). In this case, we fix the disorder realization, and set $T_1=1.5$, $T_0=7$, and disorder strength $W=8$.
Second, the PR approaches zero as $1/\mathcal{M}$ [Fig.~\ref{Fig:mbl}(b)]. 

\begin{figure}[htb]
\begin{center}
\includegraphics[width=0.6\columnwidth]{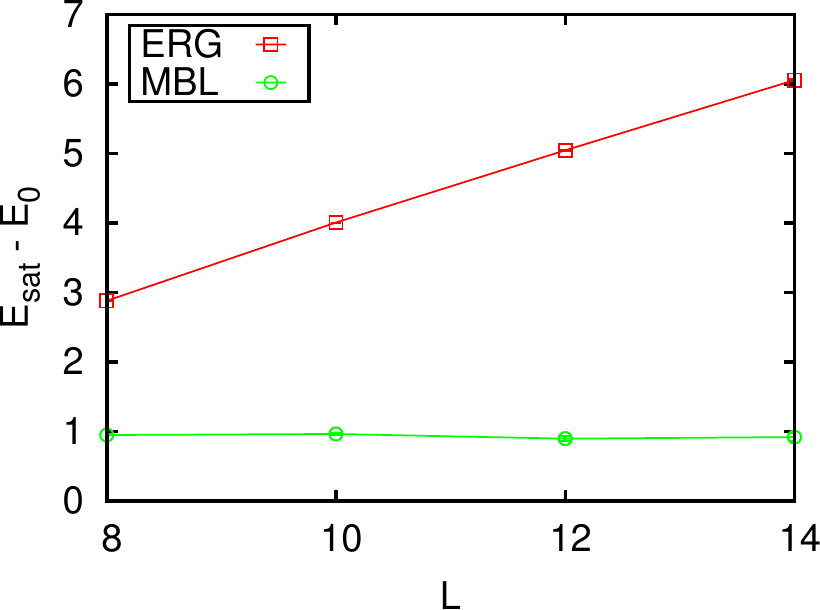}
\caption{ \label{Fig:Q} (Color online) Absorbed energy $E_{\rm sat}-E_0={\rm Tr}(\rho_\infty H_0)-E_0$ as a function of system size $L$ for ergodic (red) and MBL (green) phases. In the former case, absorbed energy is extensive and scales linearly with $L$, while in the latter case energy is absorbed only locally and its value is independent of $L$.}
\end{center}
\end{figure}

The energy absorbed after $N$ cycles, $Q_N$, when the system is initially prepared in the ground state of $H_0$, is shown in Fig.~\ref{Fig:mbl}(c). As expected for the MBL case, $Q_N$ is much smaller than 1 for all $T_1$ and decreases with system size. $Q_\infty$ remains smaller than one for all $T_1$, and decays as $1/L$ [Fig.~\ref{Fig:mbl}(d)]. These features reflect the local absorption of energy in the system. The finite-size scaling of the absorbed energy for the ergodic and MBL cases is shown in Fig.~\ref{Fig:Q}.

Finally, we have also numerically studied the level statistics of the Floquet operator, characterized by the parameter~\cite{Pal:2010gs} 
\begin{equation}
r=\langle \frac{\textrm{min}(\Delta \omega_i,\Delta \omega_{i+1})}{\textrm{max}(\Delta \omega_i,\Delta\omega_{i+1})} \rangle,
\end{equation} 
where $\Delta \omega_i = \omega_i - \omega_{i-1}$ and $\omega_i$ are chosen to lie in the interval $[0,2\pi)$. In the ergodic phase,  $r\approx 0.53$, reflecting the circular orthogonal ensemble~\cite{DAlessio:2014aa}, while in the MBL case, $r\approx 0.386$, consistent with the Poisson statistics. 

\section{Conclusions}\label{sec:conclusions}

We have shown that periodic local driving has very different effects on ergodic vs. MBL systems. Driven ergodic systems heat up to the infinite temperature, and their Floquet eigenstates are delocalized in energy space, while MBL systems absorb energy only locally, and provide an example of dynamical localization. 

A few remarks are in order. First, we expect that  our results hold for other choices of (local) $\hat V$ and $\hat H_0$, in particular for harmonic driving $\hat H(t)=\hat H_0+\hat V\cos (\Omega t)$. Second, our approach can be extended to the case of \emph{global} driving -- that is, when $\hat V$ is a sum of local terms. In this case, $\hat G$ is no longer bounded as $L\to \infty$, which can only help with delocalization~\cite{DAlessio:2014aa}. Thus, globally driven ergodic systems are also expected to heat up to the infinite temperature and to have delocalized Floquet eigenstates, in agreement with a recent study~\cite{Lazarides:2014aa}. Moreover, our results for ergodic systems suggest that non-interacting topological Floquet bands will generally be unstable to the inclusion of interactions. The case of globally driven MBL systems, on the other hand, is more intricate and deserves a separate study~\cite{Ponte:aa}.
 
Finally, we note that our results may be tested in systems of ultra-cold atoms in optical lattices~\cite{Bloch:2008ly}, trapped ions~\cite{Blatt:2012}, and NV-centers in diamond~\cite{Neumann:2008}. In systems of cold atoms, there is a potential complication of excitations into higher bands under driving; in order to observe dynamical localization or its absence in the lowest band, it is necessary to find a parameter regime where such processes are suppressed. We note that optical lattices with tunable disorder have been realized, and Anderson localization in them was observed~\cite{Kondov:2011aa}. By tuning the interactions~\cite{DeMarco:2010, Inguscio:2010, DeMarco:2013}, it should be possible to realize an MBL phase and study its response under driving. 

\section{Acknowledgements} 

We thank Luca D'Alessio, Isaac Kim, and Anatoli Polkovnikov for insightful discussions. Research at Perimeter Institute is supported by the Government of Canada through Industry Canada and by the Province of Ontario through the Ministry of Economic Development \& Innovation. D.A. is supported by NSERC Discovery grant. Z.P. acknowledges support by DOE grant DE-SC0002140. P.P. acknowledges financial support from FCT grant SFRH/BD/84875/2012. 

\section{References}

\bibliography{driven}

\end{document}